\documentclass[aps,prd,reprint,balancelastpage,nofootinbib,preprintnumbers,amsmath,amssymb,superscriptaddress,longbibliography]{revtex4-1}
\usepackage[
colorlinks=true,        % color link
citecolor=blue,         % cite color
linkcolor=blue,         % link color
urlcolor=blue           % url color
]{hyperref}             % create hyperlinks
\usepackage{bm}         % \bm{<text>} Bold math symbols
\usepackage{xcolor}     % textcolor
\usepackage{orcidlink}
\newcommand{\nc}{\newcommand*}

\nc{\Eq}[1]{Eq.~\eqref{#1}}     % equation
\nc{\Fig}[1]{Fig.~\ref{#1}}     % figure
\nc{\Table}[1]{Table~\ref{#1}}  % table
\nc{\Sec}[1]{Sec.~\ref{#1}}     % section
\def\({\left(}
\def\){\right)}
\def\[{\left[}
\def\]{\right]}
\def\e{\begin{equation}}
\def\q{\end{equation}}
\def\m{\begin{eqnarray}}
\def\n{\end{eqnarray}}

\begin{document}

\title{Pre-Big-Bang Cosmology Meets Gravitational Wave Astronomy: Constraints from  Advanced LIGO and Advanced Virgo's First Three Observing Runs}

%%%%%%%%%%%%%%%%%%%%%%%%%%%%%%%%%%%%%%%%%%%%%%%%%%%%%%%%%%%%%%%%%%%%%%%%%%%%%%%%%%%%%%%%%%%%%%%%
%%%%%%%%%%%%%%%%%%%%%%%%%%%%%%%%%%%%%%%%%%%%%%%%%%%%%%%%%%%%%%%%%
\author{Qin Tan\orcidlink{0000-0002-9496-6476}}
%\email{tanqin@hunnu.edu.cn}	
\affiliation{Department of Physics, Key Laboratory of Low Dimensional Quantum Structures and Quantum Control of Ministry of Education, Synergetic Innovation Center for Quantum Effects and Applications, Hunan Normal University, Changsha, 410081, Hunan, China}
\affiliation{Institute of Interdisciplinary Studies, Hunan Normal University, Changsha, Hunan 410081, China}

%%%%%%%%%%%%%%%%%%%%%%%%%%%%%%%%%%%%%%%%%%%%%%%%%%%%%%%%%%%%%%%%%%%%%%%%%%%%%%%%%%%%%%%%%%%%%%%%

\author{Zu-Cheng~Chen\orcidlink{0000-0001-7016-9934}}
\email{zuchengchen@hunnu.edu.cn}	
\affiliation{Department of Physics, Key Laboratory of Low Dimensional Quantum Structures and Quantum Control of Ministry of Education, Synergetic Innovation Center for Quantum Effects and Applications, Hunan Normal University, Changsha, 410081, Hunan, China}
\affiliation{Institute of Interdisciplinary Studies, Hunan Normal University, Changsha, Hunan 410081, China}

%%%%%%%%%%%%%%%%%%%%%%%%%%%%%%%%%%%%%%%%%%%%%%%%%%%%%%%%%%%%%%%%%
\author{You~Wu\orcidlink{0000-0002-9610-2284}}
%\email{Corresponding author: youwuphy@gmail.com}	
\email{youwuphy@gmail.com}
\affiliation{College of Mathematics and Physics, Hunan University of Arts and Science, Changde, 415000, China}

%%%%%%%%%%%%%%%%%%%%%%%%%%%%%%%%%%%%%%%%%%%%%%%%%%%%%%%%%%%%%%%%%
\author{Lang~Liu\orcidlink{0000-0002-0297-9633}}
%\email{Corresponding author: liulang@bnu.edu.cn}	
\email{liulang@bnu.edu.cn}	
\affiliation{Faculty of Arts and Sciences, Beijing Normal University, Zhuhai 519087, China}

%%%%%%%%%%%%%%%%%%%%%%%%%%%%%%%%%%%%%%%%%%%%%%%%
\begin{abstract}
We search for the stochastic gravitational-wave background (SGWB) predicted by pre-big-bang (PBB) cosmology using data from the first three observing runs of Advanced LIGO and Advanced Virgo. PBB cosmology proposes an alternative to cosmic inflation where the Universe evolves from a weak-coupling, low-curvature state to the hot Big Bang through a high-curvature bounce phase, predicting a distinctive SGWB spectrum. We perform a Bayesian analysis of the cross-correlation data to constrain the model parameters characterizing the PBB spectrum. We find no evidence for a PBB-induced SGWB, with a Bayes factor of $0.03$ between the PBB and noise-only model, strongly favoring the noise-only hypothesis. Our analysis establishes a lower bound $\beta \gtrsim -0.19$ at $95\%$ confidence level, which is compatible with the theoretical requirement $\beta \geq 0$ for a smooth bounce transition. While we do not detect a signal, our constraints remain consistent with the basic theoretical framework of PBB cosmology, demonstrating the potential of gravitational-wave observations to test early Universe theories.
\end{abstract}
\maketitle

%%%%%%%%%%%%%%%%%%%%%%%%%%%%%%%%%%%%%%%%%%%%%%%%%%%%%%%%%%%%%%%%%%%%%%%%%%%%%%%%%%%%%%%%%%%%%%%%
\section{Introduction}
The detection of gravitational waves (GWs) by Advanced LIGO~\cite{LIGOScientific:2014pky} and Advanced Virgo~\cite{VIRGO:2014yos} has ushered in a new era of observational astronomy~\cite{LIGOScientific:2018mvr,LIGOScientific:2020ibl,KAGRA:2021vkt}. GW observations have not only opened up a novel window to study the Universe, providing valuable insights into the physics of compact objects such as black holes and neutron stars, but have also served as a powerful tool to test the validity of general relativity in the strong-field regime. In addition to individual, high-amplitude GW events originating from merging compact binaries, the superposition of numerous weaker, unresolved GW signals can form a stochastic gravitational-wave background (SGWB). The study of SGWBs can yield crucial information about the properties and distribution of their sources, encompassing both astrophysical and cosmological origins~\cite{Regimbau:2011rp,Christensen:2018iqi}.

Astrophysical contributions to the SGWB arise from a variety of sources, including merging compact binaries and core-collapse supernovae~\cite{Zhu:2011bd,Crocker:2017agi}. On the other hand, cosmological sources are associated with various physical processes in the early Universe, such as cosmic phase transitions~\cite{Kibble:1980mv,Witten:1984rs,Mazumdar:2018dfl}, scalar-induced GWs~\cite{Ananda:2006af,Baumann:2007zm,Garcia-Bellido:2016dkw,Inomata:2016rbd,Garcia-Bellido:2017aan,Kohri:2018awv,Cai:2018dig,Papanikolaou:2024fzf}, cosmic strings~\cite{Kibble:1976sj,Sarangi:2002yt,Damour:2004kw,Siemens:2006yp}, cosmic domain walls~\cite{Vilenkin:1982ks,Sikivie:1982qv}, and primordial density perturbations during inflation~\cite{Starobinsky:1979ty,Turner:1996ck,Bar-Kana:1994nri}. These cosmological sources are isotropic and predicted to generate SGWBs with distinct spectral features, which could provide valuable insights into the physics of the early Universe. However, it is important to note that these sources are not entirely beyond the framework of general relativity, which may limit our understanding of SGWBs in the context of quantum gravity.

The primordial Universe, characterized by its extremely high energy scale, serves as a natural laboratory for studying quantum gravity. Among the existing theories of quantum gravity, string theory has garnered significant attention due to its potential to provide a unified description of all fundamental forces~\cite{Witten:1995ex,Aharony:1999ti}. String theory postulates that the fundamental building blocks of the Universe are tiny, vibrating strings of energy, which can give rise to the observed particles and forces~\cite{Schwarz:1982jn,Kaplunovsky:1985yy}. String cosmology, which applies the principles of string theory to the study of the early Universe, offers possible solutions to long-standing cosmological problems, such as the trans-Planckian problem~\cite{Martin:2000xs} and the Big Bang singularity~\cite{Borde:1993xh,Borde:2001nh}. One of the most iconic scenarios in string cosmology is the pre-big bang (PBB) scenario~\cite{Gasperini:1992em,Gasperini:1996fu,Gasperini:2007vw,Fan:2008sh,Gasperini:2016gre,Li:2019jwh,Gasperini:2021mat,Tan:2024kuk,Ben-Dayan:2024aec}, which is based on the underlying duality symmetries of string theory and has emerged as a compelling framework for generating SGWBs~\cite{Veneziano:1991ek}. This scenario predicts a cosmological phase of growing spacetime curvature and accelerated evolution, known as ``superinflation"~\cite{Lucchin:1985wy} followed by a non-singular transition to the standard radiation-dominated regime. As a result, a SGWB with a blue-tilted spectrum is naturally produced~\cite{Brustein:1995ah,Jiang:2023qht}. This distinctive spectral shape could potentially distinguish the PBB scenario from other cosmological models, such as standard slow-roll inflation~\cite{Linde:1990ta}.

Recent studies \cite{Tan:2024urn,Conzinu:2024cwl} have investigated the compatibility of the PBB scenario with the stochastic signal detected by the North American Nanohertz Observatory for Gravitational Waves (NANOGrav) and concluded that the current formulation of the PBB model cannot adequately account for the observed data. These contrasting results highlight the need for further investigation into the viability of the PBB scenario in light of the latest observational evidence. While previous studies~\cite{Tan:2024urn,Conzinu:2024cwl} focused on testing the PBB model against pulsar timing array (PTA) observations in the nanohertz frequency band, the present work extends this analysis to the significantly higher frequency band accessible to ground-based GW detectors. This multi-band approach is essential for comprehensively testing the PBB scenario, as the predicted SGWB spectrum spans many orders of magnitude in frequency, with different frequency ranges probing different epochs of the PBB evolution.

In this paper, we will use data from the first three observing runs of the LIGO-Virgo collaboration to constrain the parameters of the PBB model. Although the LIGO-Virgo collaboration has not yet detected an SGWB signal, they have determined an upper limit to its amplitude, which enables us to constrain various cosmological models. The rest of this paper is organized as follows. In Section~\ref{SGWB}, we provide an overview of the PBB scenario and its predictions for the SGWB. In Section~\ref{data}, we outline the methodology for obtaining model parameter constraints using data from Advanced LIGO and Advanced Virgo. Finally, we summarize our findings and discuss the implications of our results in Section~\ref{conclusion}.

%%%%%%%%%%%%%%%%%%%%%%%%%%%%%%%%%%%%%%%%%%%%%%%%%%%%%%%%%%%%%%%%%%%%%%%%%%%%%%%%%%%%%%%%%%%%%%%%
\section{\label{SGWB}SGWB from pre-big-bang cosmology}

In this section, we will briefly review the PBB scenario and its resulting SGWB. Due to the scale-factor duality of string cosmology~\cite{Veneziano:1991ek}, the evolution of our Universe should have a nearly mirror-symmetric phase of accelerated expansion preceding the decelerated expansion. This phase, referred to as the PBB scenario~\cite{Gasperini:1992em}, provides a possible example of primordial tensor perturbations that peak at high frequencies and exhibit a blue-tilted spectrum at low frequencies. Here we review the derivation of SGWB. The spectral energy density of the SGWB present today inside our cosmic horizon can be written as:
\begin{equation}
\Omega_{\textsc{gw}}(k,\tau_{0}) = \frac{1}{\rho_{\rm crit}(\tau_{0})} \frac{d \rho_{\textsc{gw}}}{d \ln k},\label{eq:Omega1}
\end{equation}
where $\tau_{0}$ represents the current value of the  conformal time, $\rho_{\text{crit}}=3M_{\text{Pl}}^{2}H^{2}$ is the critical energy density. Here we are concerned with the contribution to the SGWB of the cosmological amplification of perturbations of the metric tensor. For each mode $k$, the energy density is
\begin{equation}
    d\rho(\tau_{0})=2k\langle n_{k}(\tau_{0})\rangle\frac{d^{3}k}{8\pi^{3}}=\frac{k^{4}}{\pi^{2}}\langle n_{k}(\tau_{0})\rangle\ln k,\label{eq:energy density1}
\end{equation}
where $\langle n_{k}(\tau_{0})\rangle$ is the number density of gravitons produced at $\tau_{0}$. We can obtain $\langle n_{k}(\tau_{0})\rangle$ by solving the following evolution equation for the tensor mode $h_k$~\cite{Gasperini:2016gre}:
\begin{equation}
v_k'' + \left( k^2 - \frac{\xi''}{\xi} \right) v_k=0,\label{eq:evolution equation}
\end{equation}
where $v_{k}=\xi h_{k}$ and $\xi(\tau)$ is called the ``pump field''. It can be seen from the above equation that it determines the dynamics of the fluctuation $h_{k}$. For the model chosen in this paper, the background is approximated as a sequence of five cosmic phases. The pumping field $\xi$ is a simple power-law behavior like $\xi=(M_{\text{Pl}}/\sqrt{2})|\tau/\tau_{1}|^{\alpha}$ in each phase, where $\tau_{1}$ denotes the time at the end of the string phase. Then the solution $h_k$ of Eq.~\eqref{eq:evolution equation} can be expressed by the first and second kinds of Hankel functions $H^{(1)}_{\nu}$, $H^{(2)}_{\nu}$ as
\begin{equation}
    	h_k(\tau) = \left(\frac{2 \tau_1}{M_\text{Pl}^2}\right)^{\frac12} \left | \frac{\tau}{\tau_1} \right|^\nu \left[ A_+(k) H_\nu^{(2)}(k\tau)+ A_-(k) H_\nu^{(1)}(k\tau)\right].
	\label{eq:hksolution}
\end{equation}
Here, $\nu=\frac{1}{2}-\alpha$, $A_{\pm}$  are coefficients determined by the continuity of $h_k$ and $h'_k$ in each phase and by imposing the condition as $v_{k}=(1/\sqrt{2k})\exp(-ik\tau)$ for $\tau\rightarrow-\infty$. Now, the number density $\langle n_{k}(\tau_{0})\rangle$ can be expressed as
\begin{equation}
\langle n_{k}(\tau_{0})\rangle=\frac{4}{\pi}|A_{-}(k)|_{\tau=\tau_{0}}.\label{eq:number density}
\end{equation}
By combining the above equation with Eqs.~\eqref{eq:Omega1} and~\eqref{eq:energy density1}, we can get the the SGWB produced  by the PBB scenario.

\begin{figure}[tbp]
\centering
\includegraphics[width=0.35\textwidth]{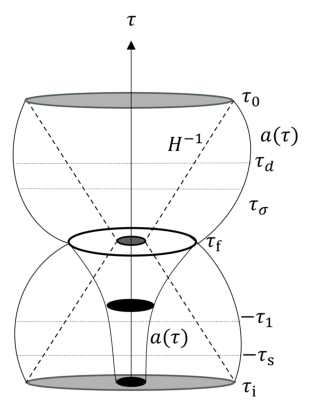}
\caption{\label{fig1} The time evolution of the Hubble horizon $H^{-1}$ (dashed line) and the scale factor $a(\tau)$ (solid line) in the PBB model. The shaded areas represent the causal connection spatial profiles of the Hubble size $H^{-1}$ at various epochs. Different $\tau$ represent the time when different phases began.}
\end{figure}

The model considered in this paper is divided into five phases by four transition  (at $\tau_{i}:\tau_1,\tau_\sigma,\tau_d,\tau_s$). Here, $\tau_1, \tau_\sigma, \tau_d, \tau_s$ correspond to the time at the end of the string phase, the beginning of a dust phase dominated by axion oscillations, the beginning of the  post Big-Bang evolution, and the moment of transition from a low energy initial stage to a possible late attractor, respectively. In Fig.~\ref{fig1}, we present the evolution of the Hubble radius $H^{-1}$ and the scale factor $a(\tau)$ in the PBB model as they evolve with conformal time. We also mark the approximate times $\tau_1,\tau_\sigma,\tau_d,\tau_s$ on the graph. In each of the above phases, the pump field has a simple power-law behavior. Specifically, The specific form of $\xi$ is~\cite{Gasperini:2016gre}
\begin{equation}
	\xi \sim \left\{
	\begin{aligned}
 \hspace{-.1cm}
  &\frac{M_\text{Pl}}{\sqrt{2}}(-\tau)^{1/2},~~~~~ \tau< -\tau_{s} \\ 
  &\frac{M_\text{Pl}}{\sqrt{2}}(-\tau)^{\beta-1},~~~  -
\tau_{s}<\tau< -\tau_{1} \\ 
  &\frac{M_\text{Pl}}{\sqrt{2}}\tau,~~~~~~~~~~~~~ -\tau_{1}<\tau< \tau_{\sigma} \\ 
  &\frac{M_\text{Pl}}{\sqrt{2}}\tau^{2},~~~~~~~~~~~~  \tau_{\sigma}<\tau< \tau_{d} \\ 
   &\frac{M_\text{Pl}}{\sqrt{2}}\tau,~~~~~~~~~~~~~ \tau_{d}<\tau< \tau_{\text{0}}.
	\end{aligned}\right. 
	\label{piecepbb}
\end{equation}
The parameter $\beta$ has deep theoretical roots in string cosmology. From the definition $\beta \equiv \frac{\mathrm{d} \log g_s}{\mathrm{d} \log a}$, it represents the rate of growth of the four-dimensional string coupling $g_s$ with respect to the scale factor $a$. This coupling evolution is crucial for the transition from the dilaton-dominated phase to the string phase and eventually to the post-big bang era. According to the specific forms of the pump field above, we can now represent the energy density fraction spectrum of SGWB as~\cite{Gasperini:2016gre,Ben-Dayan:2024aec}
\begin{equation}
 \Omega_{\text{GW}}(f) = \left\{
	\begin{aligned}
		\hspace{-.1cm}
  &\Omega_{\rm PBB}\exp{[-(f-f_1)/f_1]},~~~~   f >f_1 \\ 
   &\Omega_{\rm PBB}\left(\dfrac{f}{f_1}\right)^{ \beta_1},~~~~~~~~~~~~~~~~ f_\sigma \lesssim f \lesssim f_1 \\ 
 &\Omega_{\text{gw}}(f_1)\left(\dfrac{f_\sigma}{f_1}\right)^{\beta_1}
\left(\dfrac{f}{f_\sigma}\right)^{\beta_2}, ~ f_d \lesssim  f \lesssim f_\sigma \\ 
 &\Omega_{\text{gw}}(f_\sigma)\left(\dfrac{f_d}{f_\sigma}\right)^{\beta_2}\left(\dfrac{f}{f_d}\right)^{\beta_1},~ f_s \lesssim f \lesssim f_d\\ 
 &\Omega_{\text{gw}}(f_d)\left(\dfrac{f_s}{f_d}\right)^{\beta_1}\left(\dfrac{f}{f_s}\right)^{3}.~~~ f \lesssim f_s
	\end{aligned}\right.
	\label{eq:spectrum main}
\end{equation}
Here $f_{i}=1/(2\pi \tau_{i})$, $\beta_1=3- |3-2 \beta|$, and $\beta_2=1- |3-2 \beta|$. The dimensionless amplitude $\Omega_{\text{PBB}}$ is given by
\begin{equation}
\Omega_{\text{PBB}}=\Omega_{\text{r}0}\left(\frac{H_1}{M_{\text{Pl}}}\right)^{2} \left(\frac{f_{d}}{f_{\sigma}}\right)^{2},
\end{equation}
where  $\Omega_{\text{r}0}\approx4.15\times10^{-5}h^{-2}$ is the critical fraction of the current radiant energy density. For convenience, one can define three parameters as~\cite{Ben-Dayan:2024aec}
\begin{equation}
    z_{s}=\frac{\tau_{s}}{\tau_{1}}=\frac{f_{1}}{f_{s}},~~~~~~
z_{\sigma}=\frac{\tau_{\sigma}}{\tau_{1}}=\frac{f_{1}}{f_{\sigma}},~~~~~~
z_{d}=\frac{\tau_{d}}{\tau_{1}}=\frac{f_{1}}{f_{d}}.
\end{equation}
Now the frequencies $f_1$ and the corresponding curvature scales $H_1=H(\tau_{1})$ can be written by the above three newly defined parameters as
\begin{equation}
   f_{1}=\frac{3.9\times10^{11}}{2\pi}\left(\frac{H_{1}}{M_{\text{Pl}}}\right)^{1/2}\left(\frac{z_{\sigma}}{z_{d}}\right)^{1/2}\mathrm{Hz},\label{eq:f1}
\end{equation}
 and 
 \begin{eqnarray}
	\log_{10}\left(\frac{H_{1}}{M_{\text{Pl}}}\right)&=&\frac{2}{5 - n_{\rm s}} \left\{ \log_{10} \left[\frac{4.2 \pi^2}{T^2(H_{1})}\right] -9\right.\nonumber\\
	&&  + \left. (1-n_{\rm s})(\log_{10}  1.5 - 27)\right.\nonumber\\
	&&  + \left.(1-n_{\rm s} -2 \beta) \log_{10}  z_s \right.\nonumber\\
	&&  + \left.\frac{n_{\rm s}-1}{2} \left(\log_{10}  \frac{z_\sigma}{z_d} \right) \right\}.\label{eq:H1}
\end{eqnarray}
Here, $n_\text{s}=0.9649\pm0.0042$~\cite{Planck:2018vyg}, and
 \begin{eqnarray}
	T(H_{1})\approx &&0.13\left(\frac{H_{1}}{M_{\text{Pl}}}\right)^{1/6}z_{d}^{1/4}z_{\sigma}^{-\frac{7}{12}}\nonumber\\
	&&+0.25\left(\frac{H_{1}}{M_{\text{Pl}}}\right)^{-1/6}z_{d}^{-1/4}z_{\sigma}^{\frac{7}{12}}-0.01.\label{eq:T}     
\end{eqnarray}

Now, the spectrum of SGWB~\eqref{eq:spectrum main} is determined by only four undetermined parameters: $\beta, z_{s}, z_{\sigma}$, and $z_{d}$. It is worth pointing out that the spectrum~\eqref{eq:spectrum main} can be fitted by the following functions~\cite{Ben-Dayan:2024aec}
\begin{eqnarray}
	\Omega_{\text{GW}}(f) &=&\Omega_{\text{PBB}}f^{3}(f^{2}+f_{s}^{2})^{-\frac{|3-2\beta|}{2}}(f^{2}+f_{d}^{2})^{-1}\nonumber\\
	&&\times(f^{2}+f_{\sigma}^{2})(f^{2}+f_{1}^{2})^{\frac{|3-2\beta|-3}{2}}\nonumber \\
	&& \times\exp\left(\arctan\frac{f}{f_1}-\frac{f}{f_1}\right).\label{eq:fit}
\end{eqnarray}
It is important to note that different frequency ranges of the spectrum  probe different physical aspects of the PBB scenario. The low-frequency branch ($ f < f_s$) with its $f^3$ behavior is determined by the initial conditions and early dynamics. The intermediate branches ($f_s < f < f_1$) with their $\beta$-dependent power laws directly probe the dynamics during the high-curvature string phase. The high-frequency cutoff region ($f \sim f_1$) constrains the bounce transition scale. This multi-band structure makes it essential to test the PBB model across many orders of magnitude in frequency, from nHz (PTA) to Hz-kHz (LIGO-Virgo) bands. Figure 2 illustrates representative PBB spectra for different values of $\beta$, demonstrating how the predicted signal amplitudes compare to the current sensitivity of LIGO-Virgo detectors.
In next section, we will use data from LIGO-Virgo observations to constrain these parameters by searching for the PBB signal.

%%%%%%%%%%%%%%%%%%%%%%%%%%%%%%%%%%%%%%%%%%%%%%%%%%%%%%%%%%%%%%%%%%%%%%%%%%%%%%%%%%%%%%%%%%%%%%%%
\begin{figure}[tbp]
\centering
\includegraphics[width=0.5\textwidth]{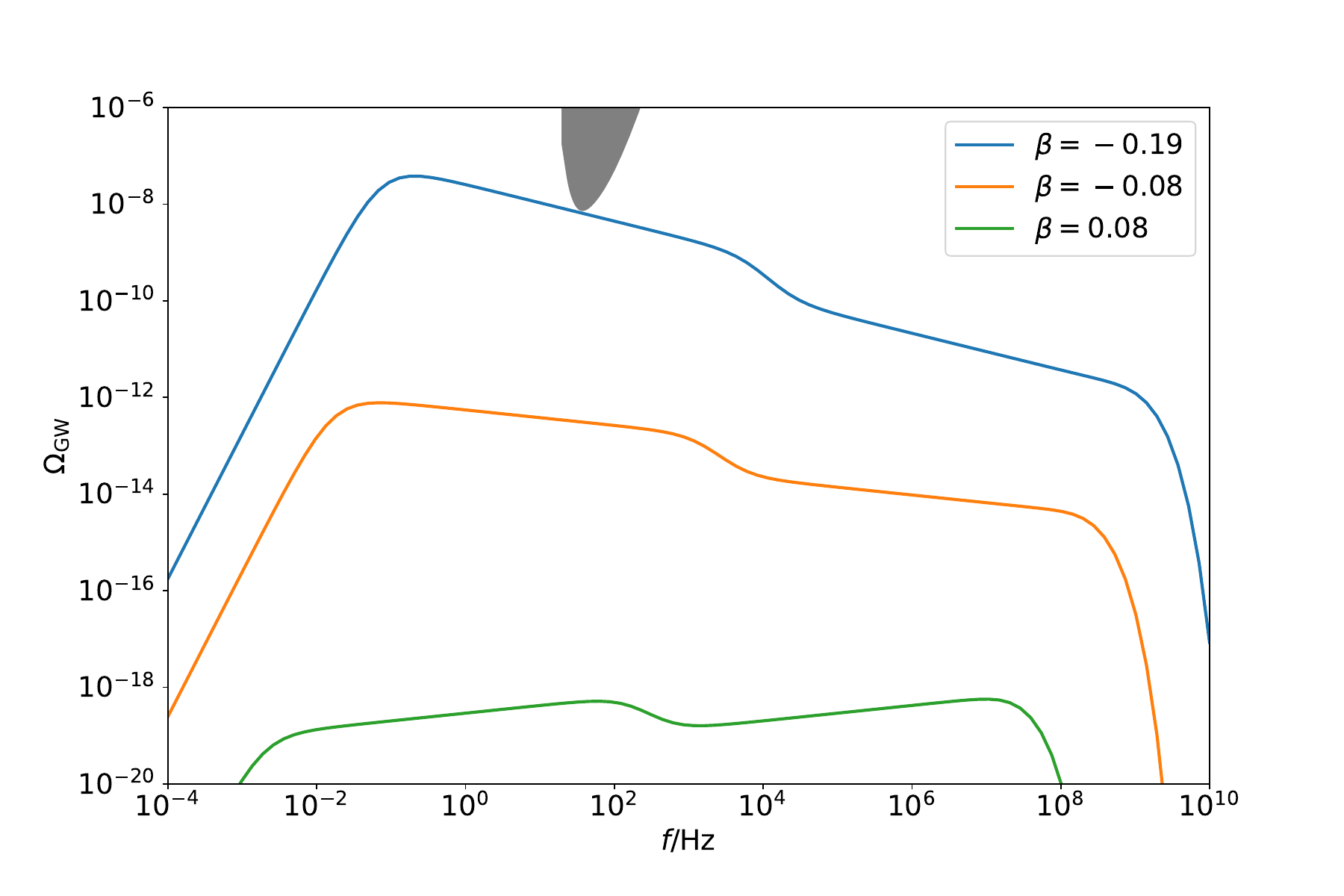}
\caption{\label{fig:ogw}Energy density spectrum of the SGWB produced by PBB scenarios compared to LVK sensitivity. The blue, orange, and green curves show representative PBB spectra with $\beta = -0.19$, $\beta = -0.08$, and $\beta = 0.08$, respectively. For all three cases, we fix the other parameters to $\log_{10} z_s = 10$, $\log_{10} z_d = 5$, and $\log_{10} z_\sigma = 4.6$. The gray shaded region represents the power-law integrated sensitivity from the LIGO-Virgo first three observing runs.}
\end{figure}
%%%%%%%%%%%%%%%%%%%%%%%%%%%%%%%%%%%%%%%%%%%%%%%%%%%%%%%%%%%%%%%%%%%%%%%%%%%%%%%%%%%%%%%%%%%%%%%%

%%%%%%%%%%%%%%%%%%%%%%%%%%%%%%%%%%%%%%%%%%%%%%%%%%%%%%%%%%%%%%%%%%%%%%%%%%%%%%%%%%%%%%%%%%%%%%%%
\section{\label{data}Data analysis}

In this section, we present the methodology employed to constrain the SGWB in the PBB model using GW data from the first three observing runs of the Advanced LIGO and Virgo detectors following our previous work \cite{Chen:2024mwg,Wu:2024deu}. The detector network consists of the LIGO-Hanford, LIGO-Livingston, and Virgo detectors, each labeled by the index $I={H, L, V}$. The analysis spans the frequency range of $20\sim 1726$~Hz, determined by the detector sensitivity and sampling rate. The time-series output, $s_I(t)$, of each detector is converted into the frequency domain using a Fourier transform, resulting in $\tilde{s}_I(f)$.

To search for the SGWB signal, we utilize the cross-correlation statistic $\hat{C}^{I J}(f)$ for each detector pair (baseline) $IJ$, as given by~\cite{Romano:2016dpx,Allen:1997ad}
\begin{equation}\label{CIJ}
\hat{C}^{I J}(f)=\frac{2}{T} \frac{\operatorname{Re}[\tilde{s}_I^{\star}(f) \tilde{s}_J(f)]}{\gamma_{I J}(f) S_0(f)},
\end{equation}
where $T$ represents the observation time, $\gamma_{I J}(f)$ denotes the normalized overlap reduction function~\cite{Allen:1997ad} that accounts for the geometric sensitivity of the detector pair, and $S_0(f)=(3 H_0^2) /(10 \pi^2 f^3)$ is a normalization factor related to the critical energy density of the Universe. The overlap reduction function is normalized such that $\gamma_{IJ}(0)=1$ for co-located and co-aligned detectors.

The cross-correlation statistic is constructed such that its expectation value equals the GW energy density spectrum, $\langle\hat{C}^{I J}(f)\rangle=\Omega_{\mathrm{GW}}(f)$, assuming no correlated noise between the detectors. For a weak SGWB signal, the variance of the cross-correlation statistic can be approximated as
\begin{equation}
\sigma_{I J}^2(f) \approx \frac{1}{2 T \Delta f} \frac{P_I(f) P_J(f)}{\gamma_{I J}^2(f) S_0^2(f)},
\end{equation}
where $P_I(f)$ denotes the one-sided power spectral density of the noise in detector $I$ and $\Delta f$ represents the frequency resolution.
The variance $\sigma_{I J}^2$ enables us to estimate the uncertainty in the cross-correlation measurement based on the detector noise properties and the observation time.

%%%%%%%%%%%%%%%%%%%%%%%%%%%%%%%%%%%%%%%%%%%%%%%%%%%%%%%%%%%%%%%%%
\begin{table}
    \centering
	\begin{tabular}{cccc}
		\hline
		Parameter & Prior & Result \\ 
  \hline \\[-2\medskipamount]   
  $\beta$ & Uniform$[-1, 3]$ & $0.42^{+1.84}_{-0.63}$\\[1pt]
  $\log_{10}z_s$ & Uniform$[0, 25]$ & $11.3^{+8.0}_{-6.6}$\\[1pt]
  $\log_{10}z_d$ & Uniform$[0, 20]$ & $6.7^{+7.8}_{-4.9}$\\[1pt]
  $\log_{10}z_\sigma$ & Uniform$[0, 18]$ & $3.7^{+7.2}_{-3.3}$\\[1pt]
  \hline
	\end{tabular}
	\caption{\label{tab:prior}Prior distributions and posterior estimates for the PBB  model parameters. The posterior estimates are reported as median values along with their corresponding $90\%$ equal-tail credible intervals.}
\end{table}
%%%%%%%%%%%%%%%%%%%%%%%%%%%%%%%%%%%%%%%%%%%%%%%%%%%%%%%%%%%%%%%%%%%%%%%%%%%%%%%%%%%%%%%%%%%%%%%%

We perform a Bayesian analysis to search for the SGWB signal originating from the PBB model, using the publicly available, model-independent cross-correlation spectra $\hat{C}^{I J}(f)$ data~\cite{KAGRA:2021kbb} from the first three observing runs of Advanced LIGO and Advanced Virgo detectors. 
To estimate the parameters of the SGWB model arising from the PBB model, we construct a likelihood function by combining the cross-correlation spectra from all detector pairs $I J$~\cite{Mandic:2012pj}:
\begin{equation}\label{like}
\scalebox{1}{$p(\hat{C}^{I J}(f_k) | \boldsymbol{\theta}) \propto \exp \left[-\frac{1}{2} \sum_{I J} \sum_k\left(\frac{\hat{C}^{I J}(f_k)-\Omega_{\mathrm{M}}(f_k | \boldsymbol{\theta})}{\sigma_{I J}^2(f_k)}\right)^2\right]$},
\end{equation}
where $\boldsymbol{\theta}$ represents the set of parameters characterizing the SGWB model, denoted by $\Omega_{\mathrm{M}}(f | \boldsymbol{\theta})$. The likelihood assumes that the cross-correlation spectra $\hat{C}^{I J}(f_k)$ follow a Gaussian distribution in the absence of a signal. The sum runs over all frequency bins $k$ and detector pairs $I J$, with $\sigma_{I J}^2(f_k)$ being the variance of the cross-correlation statistic at each frequency bin. 
Using Bayes' theorem, we express the posterior distribution of the model parameters as 
\begin{equation}
p(\boldsymbol{\theta} | C_k^{I J}) \propto p(C_k^{I J} | \boldsymbol{\theta})\, p(\boldsymbol{\theta}),
\end{equation}
where $p(\boldsymbol{\theta})$ represents the prior distribution on the parameters.

%%%%%%%%%%%%%%%%%%%%%%%%%%%%%%%%%%%%%%%%%%%%%%%%%%%%%%%%%%%%%%%%%%%%%%%%%%%%%%%%%%%%%%%%%%%%%%%%
\begin{figure}[tbp]
\centering
\includegraphics[width=0.5\textwidth]{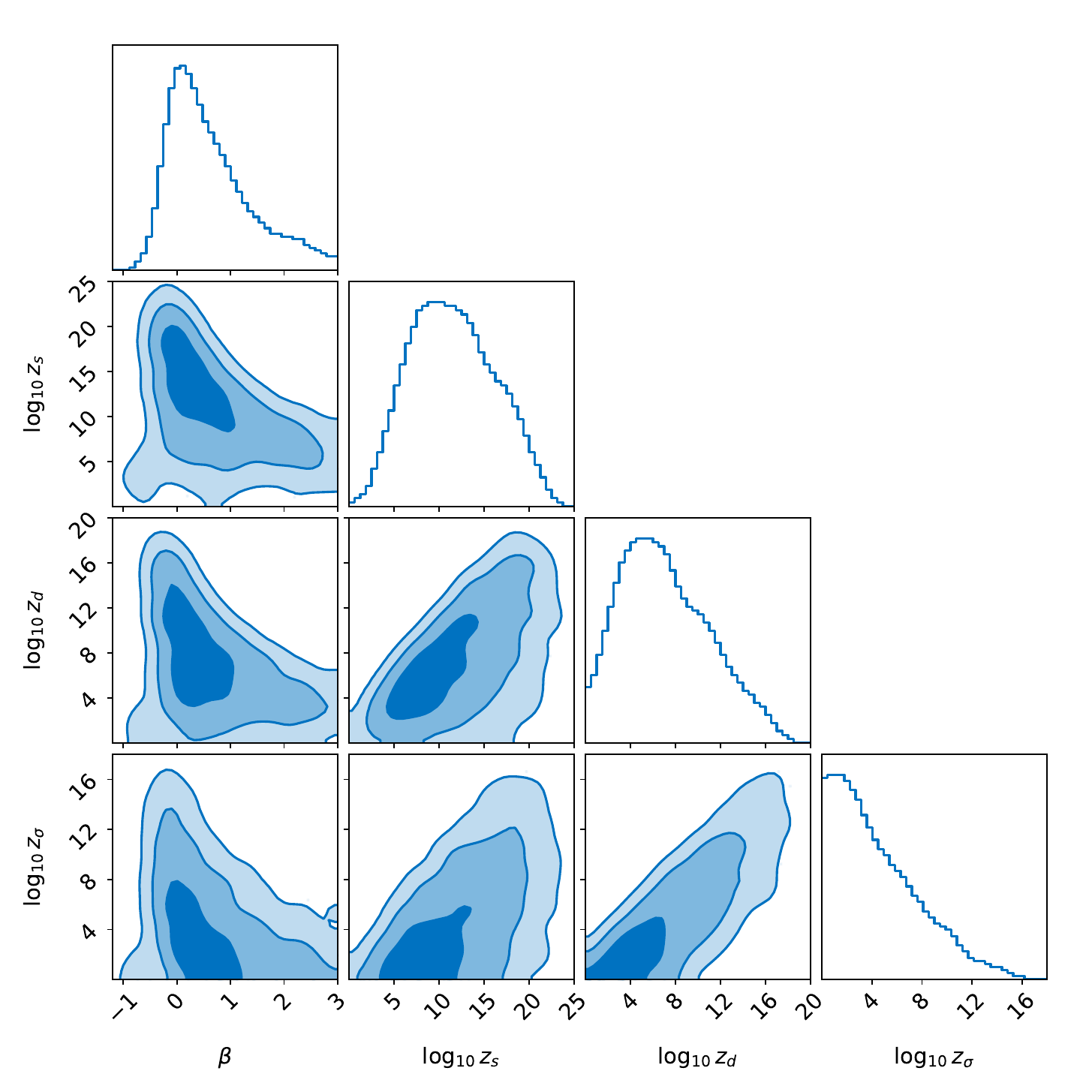}
\caption{\label{fig:posts} Posterior distributions for the PBB SGWB model parameters. The marginalized one-dimensional posteriors are shown in the diagonal panels, and the joint two-dimensional posteriors with confidence contours at $1\sigma$, $2\sigma$, and $3\sigma$ levels are displayed in the off-diagonal panels.}
\end{figure}
%%%%%%%%%%%%%%%%%%%%%%%%%%%%%%%%%%%%%%%%%%%%%%%%%%%%%%%%%%%%%%%%%%%%%%%%%%%%%%%%%%%%%%%%%%%%%%%%

To evaluate the statistical significance of the SGWB signal from the PBB model, we calculate the Bayes factor, which quantifies the relative evidence between two competing hypotheses: the model that includes the SGWB signal and the model that considers only noise,
\begin{eqnarray}
\mathcal{B}_{\mathrm{NOISE}}^{\mathrm{GW}}
&=&\frac{p(\hat{C}^{IJ} | \text{Model with SGWB signal})}{p(\hat{C}^{IJ} | \text{Pure noise model})} \nonumber\\
&=& \frac{\int p(\hat{C}^{IJ}| \boldsymbol{\theta}_{\mathrm{GW}})\, p(\boldsymbol{\theta}_{\mathrm{GW}})\, \mathrm{d} \boldsymbol{\theta}_{\mathrm{GW}}}{\mathcal{N}}.
\end{eqnarray}
The numerator denotes the marginal likelihood of the model incorporating the SGWB signal, which is calculated by integrating the likelihood $p(\hat{C}^{IJ}| \boldsymbol{\theta}_{\mathrm{GW}})$ multiplied by the prior $p(\boldsymbol{\theta}_{\mathrm{GW}})$ across the parameter space $\boldsymbol{\theta}_{\mathrm{GW}}$. The denominator $\mathcal{N}$ represents the evidence for the pure noise model, obtained by setting $\Omega_{\mathrm{M}}(f)=0$ in Eq.~\eqref{like}. It represents the probability of observing the data given the model that assumes only the presence of noise.
The Bayes factor provides a quantitative measure of the relative support for the model with the SGWB signal compared to the pure noise model. A value of $\mathcal{B}_{\mathrm{NOISE}}^{\mathrm{GW}}>1$ indicates that the data favor the model with the SGWB signal over the pure noise model. The strength of the evidence can be interpreted using a standard scale, such as the Jeffreys scale~\cite{Jeffreys:1939xee}, where Bayes factors $\mathcal{B}_{\mathrm{NOISE}}^{\mathrm{GW}}$ exceeding 3, 10, 30, and 100 indicate substantial, strong, very strong, and decisive evidence in favor of the model with the SGWB signal, respectively.  

The free parameters in our analysis are $\boldsymbol{\theta}_{\mathrm{GW}} \equiv (\beta, z_s, z_d, z_\sigma)$. We list the priors of the four free parameters in Table~\ref{tab:prior}. These parameters are subject to the following theoretical constraints~\cite{Ben-Dayan:2024aec}: 
\begin{align}
	&1 \lesssim z_\sigma<z_d<z_s,\\
	&\log _{10}\left(\frac{H_1}{M_{\mathrm{Pl}}}\right)+\frac{3}{2} \log _{10} z_d-\frac{7}{2} \log _{10} z_\sigma \leqslant 0,\\
	&\log _{10}\left(\frac{H_1}{M_{\mathrm{Pl}}}\right)-3 \log _{10} z_d+\log _{10} z_\sigma>-42-\log _{10} 4,\\
	&\log _{10} z_s<26-\log _{10} 9+\frac{1}{2} \log _{10}\left(\frac{H_1}{M_{\mathrm{Pl}}}\right)\nonumber\\
	&\qquad \qquad \qquad +\frac{1}{2}\left(\log _{10} z_\sigma-\log _{10} z_d\right).
\end{align}
We conduct the Bayesian analysis using the \texttt{Bilby} package~\cite{Ashton:2018jfp,Romero-Shaw:2020owr}, employing the dynamic nested sampling algorithm implemented in \texttt{Dynesty}~\cite{Speagle:2019ivv} with $1024$ live points to ensure adequate sampling of the parameter space.

%%%%%%%%%%%%%%%%%%%%%%%%%%%%%%%%%%%%%%%%%%%%%%%%%%%%%%%%%%%%%%%%%%%%%%%%%%%%%%%%%%%%%%%%%%%%%%%%
\section{\label{conclusion}Results and discussions}

PBB cosmological models offer an alternative to the standard inflationary paradigm, proposing that the Universe existed in a low-energy string phase prior to the Big Bang. One of the key predictions of PBB models is the generation of a SGWB with a distinctive spectrum. In this study, we search for the SGWB signal predicted by PBB cosmology using data from the first three observing runs of Advanced LIGO and Advanced Virgo. By parameterizing the predicted SGWB spectrum and performing a Bayesian analysis, we constrain the model parameters, obtaining $\beta = 0.42^{+1.84}_{-0.63}$, $\log_{10} z_s = 11.3^{+8.0}_{-6.6}$, $\log_{10}z_d=6.7^{+7.8}_{-4.9}$, and $\log_{10}z_\sigma=3.7^{+7.2}_{-3.3}$. The posterior distributions for the parameters of the SGWB model from the PBB cosmology are presented in Figure \ref{fig:posts}.

Our analysis reveals no statistically significant evidence for the presence of a PBB SGWB signal in the Advanced LIGO and Advanced Virgo data. The Bayes factor between the PBB model and the noise-only model is found to be $0.03$, indicating ``very strong" preference for the noise-only model. Consequently, we establish a lower limit for the parameter $\beta$ as $\beta \gtrsim -0.19$ at 90\% confidence level. 
The theoretically viable range for this parameter is $0 \leq \beta < 3$, where the lower bound arises from the requirement of growing string coupling necessary for a smooth bounce transition~\cite{Gasperini:2023tus,Conzinu:2023fth,Modesto:2022asj}, and the upper bound prevents background instabilities~\cite{Kawai:1998ab}.
Notably, our observational lower limit of $\beta \gtrsim -0.19$ is compatible with the theoretical constraint $\beta \geq 0$, indicating that while we do not detect a SGWB signal, our results remain consistent with the basic theoretical framework of PBB cosmology.

Our constraint $\beta \gtrsim -0.19$ at 95\% C.L., while less stringent than the theoretical requirement $\beta \gtrsim 0$, provides crucial independent validation of the PBB framework. This experimental bound serves multiple important purposes: First, it offers a model-agnostic test of theoretical predictions based on string theory and stability requirements. Second, the consistency between our experimental constraint and theoretical expectations supports the validity of the underlying PBB framework—had we found $\beta \ll -0.19$, this would have challenged the entire theoretical structure. Third, our constraint from the Hz-kHz band complements and strengthens the tension revealed by PTA analyses \cite{Tan:2024urn,Conzinu:2024cwl}, which require $\beta \approx -0.12$ to fit NANOGrav data, thereby violating theoretical bounds. This multi-band consistency demonstrates that strongly negative $\beta$ values are disfavored across different frequency ranges.

Our findings underscore the potential of GW experiments in probing alternative cosmological models and exploring the pre-big-bang era. As the sensitivity of GW detectors continues to improve, we anticipate placing even more stringent constraints on PBB models or potentially detecting the PBB SGWB signal. The upcoming observing runs of Advanced LIGO and Advanced Virgo, along with future detectors such as the Einstein Telescope~\cite{Punturo:2010zz} and Cosmic Explorer~\cite{Reitze:2019iox}, will play pivotal roles in this endeavor, opening new avenues for investigating the earliest stages of the Universe and testing the fundamental principles of string theory and quantum gravity.

%%%%%%%%%%%%%%%%%%%%%%%%%%%%%%%%%%%%%%%%%%%%%%%%%%%%%%%%%%%%%%%%%
\begin{acknowledgements}
QT is supported by the National Natural Science Foundation of China (Grants No.~12405055 and No.~12347111), the China Postdoctoral Science Foundation (Grant No.~2023M741148), the Postdoctoral Fellowship Program of CPSF (Grant No. GZC20240458).
ZCC is supported by the National Natural Science Foundation of China under Grant No.~12405056, the Natural Science Foundation of Hunan Province under Grant No.~2025JJ40006, and the Innovative Research Group of Hunan Province under Grant No.~2024JJ1006. 
YW is supported by the National Natural Science Foundation of China under Grant No.~12405057.
LL is supported by the National Natural Science Foundation of China (Grant No.~12505054 and ~12433001).
\end{acknowledgements}

\bibliography{ref}
\end{document}